\documentclass[number,sort&compress]{elsarticle}

\usepackage{graphicx}
\usepackage{amsmath}   
\usepackage{upgreek}
\usepackage{amssymb}
\usepackage{color}
\usepackage{url}
\usepackage[binary-units=true]{siunitx}
\usepackage{lmodern}

\begin{document}

\begin{frontmatter}

\title{%
Status of the vertex detector program of the CBM experiment at FAIR}

\author{Philipp Klaus\corref{cor1}}
\ead{klaus@physik.uni-frankfurt.de}
\cortext[cor1]{Corresponding author: Philipp Klaus}

\author{Michal Koziel, Ole Artz, Norbert Bialas, Michael Deveaux, Ingo Fr\"{o}hlich, Jan Michel, Christian M\"{u}ntz, Roland Weirich, Joachim Stroth}
\address{Goethe-Universit\"{a}t Frankfurt, Institut f\"{u}r Kernphysik, Max-von-Laue Strasse 1, 60438 Frankfurt am Main, Germany.}

\begin{abstract}
The Compressed Baryonic Matter Experiment (CBM) is one of the core experiments of the future FAIR facility (Darmstadt/Germany). The fixed-target experiment will explore the phase diagram of strongly interacting matter in the regime of high net baryon densities with numerous rare probes. The Micro Vertex Detector (MVD) will determine the secondary decay vertex of open charm particles with $\sim 50~\rm \upmu m$ precision, contribute to the background rejection in dielectron spectroscopy, and help to reconstruct neutral decay products of strange particles by means of missing mass identification. The MVD comprises four stations with 0.3 and $0.5\%~x/X_0$, which are placed between 5 and $20~\rm cm$ downstream the target and inside vacuum. It will host highly-granular, next-generation Monolithic Active Pixel Sensors, with a spatial precision of $5\rm~\upmu m$, a time resolution of $5 ~\rm \upmu s$, and a peak rate capability of $\sim 700~\rm kHz/mm^2$. Moreover, a tolerance to $3\cdot10^{13}~\rm n_{eq}/cm^{2}$ and $\gtrsim 3~\rm Mrad$ are required.
In this document, we summarize the status of sensor development, station prototyping, and the detector slow control.

\end{abstract}


\end{frontmatter}

\section{Introduction}
The MVD of the fixed target Compressed Baryonic Matter Experiment \cite{CBM} will consist of four CMOS pixel detector stations \cite{CPS} populated with radiation-tolerant sensors, form-factor $31\times\SI{17}{\milli\meter^2}$, thinned to \SI{50}{\um}.
An ultra-light and vacuum-compatible cooling concept is foreseen to transfer the dissipated power of the sensors via highly heat-conductive carriers (CVD-diamond \cite{CVD} and Thermal Pyrolytic Graphite \cite{TPG}), in turn clamped into liquid cooled aluminum-based heat sinks mounted outside the detector acceptance. Control, power, and data links are provided to the sensors via customized, thin Flexible Printed Circuit (FPC) cables. The sensors exposed to the maximum hit densities will be read out by the mean of eight differential data links.
This paper summarizes the status of sensor development, station prototyping, and detector slow control.

\section{Sensor Development}

\begin{figure}[h]
	\centering
	\includegraphics[width=8cm,keepaspectratio, bb= 1 4 600 400]{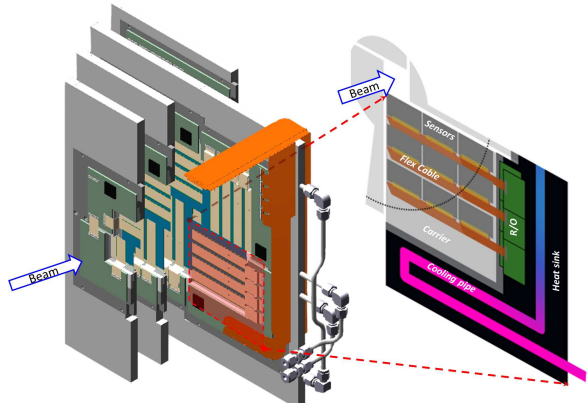} 
	\caption{Visualization of the Micro-Vertex Detector of the CBM experiment. Station three is fully depicted. It employs nine sensors on one side and six on the other side, respectively. The TPG carrier is clamped into an aluminum heat sink that holds also passive R/O electronics (DC filters and impedance matching).}
	\label{fig:MVD}
\end{figure}

There is no ready technical solution that could provide a CMOS Pixel Sensor (CPS) suitable for the CBM experiment, e.g.~the ALICE-ALPIDE chip~\cite{ALPIDE} is not adapted to the maximum hit rate and radiation load at CBM (driven by local hot-spots due to $\delta$-electrons in the fixed-target setup). The CBM-MVD sensor, called "MIMOSIS", is being developed by IPHC Strasbourg with the TowerJazz 0.18~$\si{micrometer}$ process, same as used for manufacturing the ALPIDE chip. MIMOSIS will profit of ALPIDE's readout but feature an in-pixel architecture more tolerant to radiation and will be equipped with entirely new digital micro-circuitry (chip communication, signal processing and control). The sensor development is made in several steps, where the first is realized with a prototype chip called MIMOSIS-0, aiming at selecting an optimum in-pixel architecture and studying the built-in priority encoder. 
To characterize the sensor performance, a dedicated readout system that provides sensor powering and data transmission from the sensor to the computer, has been developed. To speed up test campaign, several test benches were set up and distributed within the participating institutes. The planned radiation tolerance tests will cover the doses up to 10~$\si{Mrad}$ and $10^{14}\si{n_{eq}/\cm^2}$. We focus on selecting the most robust in-pixel architecture and characterizing, after irradiation, the performance of digital to analog converters used to define reference voltages and operating points for, e.g.~in-pixel amplifiers and discriminators.


\section{Prototyping}
There were two prototyping phases accomplished validating the feasibility of use the CVD diamond and TPG materials as a carrier capable to most efficiently transfer the heat from the MVD sensors. The prototype described in \cite{MVDprototype} was based on the CVD-diamond and hosting two 50~$\si{\micro\meter}$-thin \mbox{MIMOSA-26} sensors~\cite{FirstRetMAPS,M26AHR} on both sides of the CVD carrier. The second prototype, called "PRESTO"~\cite{PRESTO}, comprises 15 thinned CPS of the same type. Nine of the sensors were glued on the front, and six on the back side of the module. Figure \ref{fig:MVD} illustrates the full detector and PRESTO. The sensors are wire bonded to a total of 10 FPC cables, which provide the necessary bias lines and data links. PRESTO has the size of a quadrant of the 2nd MVD station but due to the number of sensors integrated, its complexity is equivalent to the one of the 3rd station. The assembly procedures and QA measures were established, guaranteeing vacuum compatible integration of the sensors on both sides of the carrier with a placing precision of better than \SI{100}{\um}. Cooling concept, selection of adhesives, vacuum compatibility, and FPC cable performance \cite{ROchain} were studied with a positive outcome. The rather poor sensor assembly yield of $\sim60\%$ for the PRESTO module was identified as problems with a wire bonding machine. This will be proven by a follow-up prototype.

\section{Detector Slow Control}
For 24/7 reliability tests of the PRESTO module and to ensure a failsafe operation of the device at any time, the EPICS framework (v3.15.5)~\cite{epics} is used to monitor and control the system. We have implemented the required I/O Controllers (IOCs), those are the EPICS server processes that establish the connection between the laboratory equipment and the EPICS network called channel access (CA). The CS-Studio RDB Archiver was chosen to store process variables in a PostgreSQL database set up with table partitioning, see~\cite{b}. What concerns the user interfaces, a two-fold strategy was selected: A {CS-Studio interface was created for full-fledged system control, locally accessible for the operators and experts. In addition, a highly configurable web dashboard for EPICS PVs was developed. Running in any modern browser, it can be used on any device with an internet connection. It was designed to be r/o. In case user logins and protocolling would be added later, write acces could be considered. At the moment, we are implementing cross-IOC alarm handling and procedures to startup\,/\,shutdown the detector. This work serves also as a small size prototype of the CBM-MVD slow control system.
	

\section*{Acknowledgments}
We would like to thank the IPHC-PICSEL team for their work. Work supported by BMBF (05P15RFFC1), HIC for FAIR and GSI.


\end{document}